\documentclass[aps,prb,reprint,twocolumn,showpacs]{revtex4-1}

\usepackage[ansinew]{inputenc}
\usepackage[T1]{fontenc}
\usepackage{ae,aecompl}
\usepackage{graphicx}
\usepackage{epstopdf}
\usepackage{color}

\begin{document}

\title{Superconducting Sweet-Spot in Microcrystalline Graphite Revealed by Point-Contact Spectroscopy }

\author{F. Arnold, J. Ny\'eki, J. Saunders\/\thanks{correspondence should be addressed to: j.saunders@rhul.ac.uk}}
\affiliation{Royal Holloway, University of London, Egham Hill, TW20 0EX, Egham, United Kingdom}

\date{30 Mach 2018}


\begin{abstract}In this letter we describe the observation of a magnetic field dependent electronic gap, suggestive of local superconductivity, in the point-contact spectrum of micro-crystalline graphite. Magnetic field dependent point-contact spectroscopy was carried out at a temperature of $1.8\,\mathrm{K}$ using an etched aluminium tip. At zero field a gap structure in the differential conductance is observed, showing a gap of $\Delta = 4.2\,\mathrm{meV}$. On applying magnetic fields of up to $500\,\mathrm{mT}$, this gap gradually closes, following the theoretical prediction by Ginzburg and Landau for a fully flux-penetrated superconductor. By applying BCS-theory, we infer a critical superconducting temperature of $14\,\mathrm{K}$. 
\end{abstract}

\maketitle

The possibility of high temperature superconductivity in bulk graphite-based materials remains an open question. A number of intriguing observations have been reported, based on studies of magnetization and electrical transport. These are interpreted in terms of superconducting regions, surviving to above room temperature, localized at internal interfaces within the sample \cite{Esquinazi18}. Elsewhere, it has been shown theoretically that flat bands at certain interfaces, for example between inclusions of rhombohedral graphite and stable Bernal graphite, strongly favor superconductivity \cite{Volovik18,Kopnin11}. On the other hand experimental evidence for superconductivity in a twisted graphene bilayer has been recently reported. A broad transition in the resistance is observed, with onset around $1.7\,\mathrm{K}$ and zero resistance within experimental resolution at $70\,\mathrm{mK}$ \cite{Cao18_26160}. This emphasizes the importance of defects, interfaces and other deviations from the ideal Bernal graphite structure for the occurrence of superconductivity in graphite based materials.

In this article, we show the emergence of a superconducting state with a $T_\mathrm{c}$ of $14\,\mathrm{K}$ in micro-crystalline graphite observed by point-contact spectroscopy. Point-contact spectroscopy is a powerful technique to probe the local density-of-states and electronic spectrum of a metal \cite{Yanson81,Jansen80}. Micro-crystalline graphite, Grafoil \cite{UCAR}, is prepared by thermochemical exfoliation and subsequent recompression, leading to foils of interlinked co-aligned graphite micro-crystallites.  Point-contact spectra were measured in a Quantum Design PPMS using a home made spectrometer \cite{Arnold15Thesis}. By fine tuning the needle position and using electrochemically etched aluminium tips with typical tip radii of a few micrometers, it was possible to alter the contact resistance over three orders of magnitude to ensure a ballistic contact regime. The general point-contact spectra of graphite follow an almost symmetrical v-shape centered at zero-bias. This v-shape arises due to the semimetallic density-of-states of graphite \cite{McClure57}.

\begin{figure}[tb]
	\centering
		\includegraphics[width=1\columnwidth]{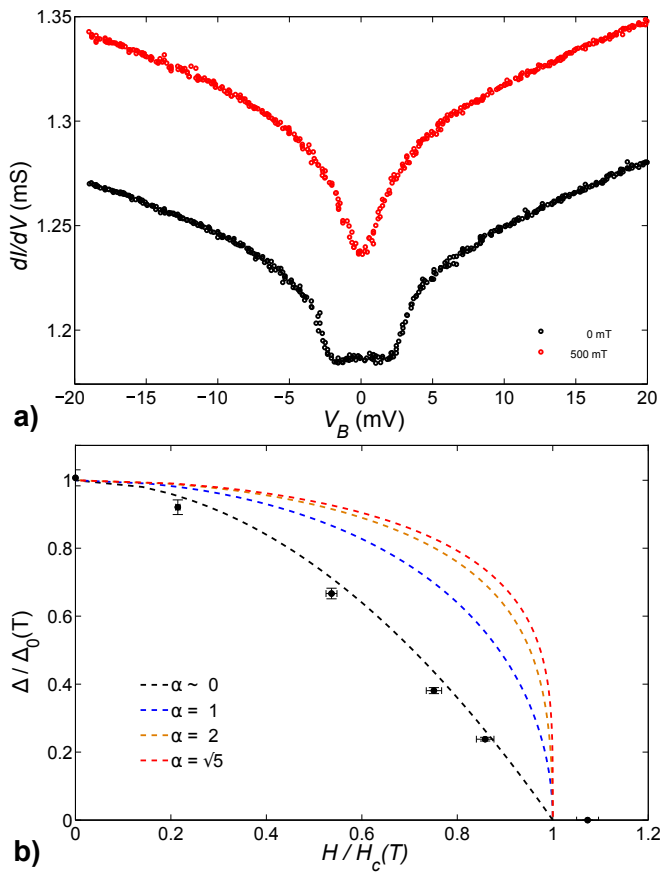}
	\caption{Fig. 1: {\bf a)} shows the point-contact spectra of the superconducting sweet-spot at $1.8\,\mathrm{K}$ in zero and $500\,\mathrm{mT}$ field {\bf b)} the magnetic field dependence of the inferred gap $\Delta$ renormalized by its zero-field value $\Delta_0(T)$ and critical field $\mu_0H_c=450\,\mathrm{mT}$. The dashed lines are theoretical predictions for BCS-superconductors in the limit of a fully field penetrated superconductor \cite{Douglass61,Meservey64}.}
	\label{fig:Figure1}
\end{figure}

During measurements on micro-crystalline graphite, we serendipitously found an "`anomalous"' point-contact spectrum (see Fig. \ref{fig:Figure1}a)). At large bias voltages the point-contact spectrum resembles the v-shape observed in bulk graphite. However, at lower bias a plateau appears, spanning $\Delta_0 = 4.2\,\mathrm{meV}$ at zero field. Such a plateau is  attributed to a gap in the electronic spectrum, where the density-of-states goes to zero \cite{Blonder82}. On applying a magnetic field this plateau is gradually suppressed and vanishing at $450\,\mathrm{mT}$. In Fig. \ref{fig:Figure1}b) the magnetic field dependence of the inferred gap size is shown.

As proposed in \cite{Arnold15Thesis}, the magnetic field dependence of this putative gap is consistent with the suppression of a superconducting gap. The experimental gap (Fig. \ref{fig:Figure1}) follows the magnetic field dependence of a fully field penetrated BCS-superconductor, where the superconducting domains are much smaller than the London penetration depth \cite{Douglass61,Meservey64,London48}. The theoretical magnetic field dependencies of field penetrated superconductors are plotted for different ratios of the superconducting domain size $d$ to the London penetration depth $\lambda$. As the magnetic field penetrates a superconductor only on the length scale of $\lambda$ these effects become prominent when the superconducting domain size and London penetration depth are of equal size. As can be seen our data is best fit by the theoretical curve for $d/\lambda\approx0$, which corresponds to almost complete suppression of the Mei\ss ner effect. Assuming the observed state is a BCS-superconductor, its critical temperature can be estimated by applying the BCS-formula, $\Delta_0=1.764k_\mathrm{B}T_c$, to the measured zero-field gap. Using $\Delta_0 = 4.2\,\mathrm{meV}$, we find that the critical temperature of the superconducting state $T_c \approx 14\,\mathrm{K}$. This is significantly larger than the value inferred from twisted bilayer graphene \cite{Cao18_26160}.

In contrast to other point-contact spectra on BCS-superconductors, we do not observe Andreev reflections within the superconducting gap, proving the presence of Cooper pairs \cite{Andreev64}. The possibility of strongly enhanced superconductivity in mesoscopic aluminium should also be considered. However, the observed critical temperature and gap size is seven times larger than the highest observed superconducting gaps in mesoscopic aluminium ($\Delta\approx300\,\mu\mathrm{eV}$) \cite{Cohen68,Black96,Court08}. It disappears when the needle is moved to another position. Therefore, we exclude the possibility of aluminium superconductivity.

Crucially in our experiment, Grafoil is a highly inhomogeneous graphite allotrope, with a large number of crystal defects. We propose that our measurement serendipitously revealed a so far unknown crystal defect, which enables so far unseen high temperature superconductivity in graphite. Systematic investigations to isolate and characterize the microstructure within graphite responsible for superconductivity are highly desirable, to complement "`bottom-up"' studies using graphene. If room temperature superconductivity is indeed accessible via this common allotrope of carbon, the potential impact is significant.\\
The authors wish to thank the Engineering and Physical Science Research Council (EPSRC Grant Nos.  EP/H048375/1 and EP/J010618/1) and European Microkelvin Platform for financial support as well as B. Cowan, A. Casey, L.V. Levitin, Ch. Howard, B. Yager and G. E. Volovik for fruitful discussions.


\begin{thebibliography}{18}

\bibitem{Esquinazi18}
for the most recent review see P.~D. Esquinazi {\it et al.}, Quantum Stud.: Math. Found.,{ \bf 5}, 41 (2018) or arXiv 1709.00259 and references therein.

\bibitem{Volovik18}
G.~E. Volovik, Pis'ma v ZhETF,{ \bf } (2018).

\bibitem{Kopnin11}
N. Kopnin, T. Heikkilae, and G.~E. Volovik, Phys. Rev. B,{ \bf 83}, (220503(R)) (2011).

\bibitem{Cao18_26160}
Y. Cao {\it et al.} {\it Unconventional superconductivity in magic-angle graphene
  superlattices.}, Nature, http://dx.doi.org/10.1038/nature26160 (2018).

\bibitem{Yanson81}
I.~K. Yanson, I.~O. Kulik and A.~G. Batrak, J. Low Temp. Phys.,{ \bf 42}(5-6), 527--556 (1981).

\bibitem{Jansen80}
A.~G.~M. Jansen, A.~P. van Gelder and P. Wyder, J. Phys. C: Solid St. Phys.,{ \bf 13}, 6073--6118 (1980).

\bibitem{Lee14}
W.-C. Lee, W. Park, H. Arham, L. Greene and P. Phillips, PNAS,{ \bf 112}(3), 651--656 (2015).

\bibitem{UCAR}
Product of GrafTech Internat. Adv. Elec. Tech., http://www.graftech.com.

\bibitem{Arnold15Thesis}
F. Arnold {\it Experimental Study of Strongly Correlated Fermion Systems under
  Extreme Conditions: Two-Dimensional $^3$He at Ultra-Low Temperatures and
  Graphite in the Magnetic Ultra-Quantum Limit}. Thesis, Royal Holloway, University of London, (2015).

\bibitem{McClure57}
J.~W. McClure, Phys. Rev.,{ \bf 108}(3), 612--618 (1957).

\bibitem{Blonder82}
G.~E.Blonder, M.Tinkham, and T.~M. Klapwijk, Phys. Rev. B,{ \bf 25}, 4515 (1982).

\bibitem{Douglass61}
D.~H. Douglass Jr., Phys. Rev. Lett.,{ \bf 6}(7), 346--348 (1961) .

\bibitem{Meservey64}
R. Meservey and D. H. Douglas Jr., Phys. Rev.,{ \bf 135}(1A), A24--A33 (1964).

\bibitem{London48}
F. London, Phys. Rev.,{ \bf 74}(4), 562--573 (1948).

\bibitem{Andreev64}
A. Andreev, Sov. Phys. JETP,{ \bf 19}(5), 1228--1231 (1964).

\bibitem{Cohen68}
R. Cohen and B. Abeles, Phys. Rev.,{ \bf 168}(2), 444--450 (1968).

\bibitem{Black96}
C. Black, D. Ralph and M. Tinkham, Phys. Rev. Lett.,{ \bf 76}(4), 688--691 (1996).

\bibitem{Court08}
N. Court, A. Ferguson and R. Clark, Supercond. Sci. Technol.,{ \bf 21}, (015013) (2008).

\end{thebibliography}
\end{document}